\documentclass[paper]{revtex4}

\usepackage{amsmath,amssymb,mathrsfs,bbm}
\usepackage{graphicx}
\usepackage{amsthm}
\usepackage{amscd}
\usepackage{bm}
\usepackage{dsfont}
\usepackage[utf8]{inputenc}

\begin{document}

\title{Superposition principle and Kirchhoff's integral theorem}
\author{M. I. Krivoruchenko}
\affiliation{
Institute for Theoretical and Experimental Physics$\mathrm{,}$ B. Cheremushkinskaya 25\\
117218 Moscow, Russia}
\affiliation{
Moscow Institute of Physics and Technology$\mathrm{,}$ 141700 Dolgoprudny$\mathrm{,}$ Russia}
\affiliation{
Bogoliubov Laboratory of Theoretical Physics$\mathrm{,}$ Joint Institute for Nuclear Research\\
141980 Dubna$\mathrm{,}$ Russia
}


\begin{abstract}
The need for modification of the Huygens-Fresnel superposition principle arises even
in the description of the free fields of massive particles
and, more extensively,
in nonlinear field theories. A wide range of formulations and superposition schemes for secondary waves are captured by Kirchhoff's integral theorem.
We discuss various versions of this 
theorem
as well as
its connection with the superposition principle and the method of Green's functions.
A superposition scheme inherent in linear field theories, which is not based on Kirchhoff's integral theorem but instead relies on the completeness condition, is also discussed.
\end{abstract}

\maketitle

\setcounter{page}{1}

\pagebreak
\baselineskip= 14pt
\tableofcontents
\newpage

\baselineskip= 15pt

\section{Introduction}
\renewcommand{\theequation}{I.\arabic{equation}}
\setcounter{equation}{0}

An excellent and detailed explanation of Huygens' principle for undergraduate students, together with the optical-mechanical analogy and the Hamilton-Jacobi method, can be found in the monograph by Arnold \cite{Arno89}. Students are introduced to a generalization of Huygens' principle, viz. the Huygens-Fresnel superposition principle, in the study of general physics (see, e.g., \cite{Save82}), and this principle is presented in greater detail in the study of theoretical physics (see, e.g., \cite{Land71}). The method of Green's functions (GF)
which has found numerous applications in a large variety of different fields
is discussed in the first volume of a two-volume monograph by Bjorken and Drell \cite{Bjor64,Bjor65}, where, in particular, the superposition principle is used in~\S\S~21~-~22 to derive the equation for the Green's function.
Further development of concepts related to the superposition principle 
has led to the emergence in quantum theory of the path integral formalism,
an excellent overview of which can be found in the monograph by Dittrich and Reuter \cite{DITT2001}. This material is intended for advanced students studying quantum field theory.
A detailed presentation of the superposition principle for electromagnetic fields, its rationale and its generalizations, based on Kirchhoff's integral theorem,
\cite{Kirch1883}
is given in the monograph by Born and Wolf \cite{Born99}. This monograph is intended for postgraduate students and researchers specializing in the theory of the propagation of electromagnetic waves and light phenomena.

Thus, it is clear that the superposition principle is closely related to the GF method which, in turn, lies at the heart of quantum field theory and the diagram technique. In the literature, this relationship is typically mentioned only in passing, while the mathematical aspects, modifications, and physical meaning of the generalized schemes of superposition are treated as matters beyond dispute.

A rigorous 
formulation of the
superposition principle is based on Kirchhoff's integral theorem. The generalizations to which it leads are used also in the theory of interacting fields.
In this paper, we attempt to specify the precise place of the superposition principle in classical and quantum field theory and discuss its relationship with the GF method and Kirchhoff's integral theorem.

Surprisingly, the answers to the main questions can be obtained by analyzing the dynamics of the one-dimensional oscillator. The oscillator problem from the viewpoint of Kirchhoff's integral theorem, as well as its connections with the superposition principle and the GF method, is discussed in the next section.
In Sect. III, we consider a free massive scalar field. For massive fields, the superposition scheme includes an integral over three-dimensional space. Both in the limit of zero mass and for monochromatic fields, the canonical superposition scheme, in which the summation of the sources of secondary waves is limited to a two-dimensional surface, arises.
The statement of Kirchhoff's theorem depends on the asymptotic conditions imposed on the propagator at $t\to \pm \infty $. In quantum field theory, the Feynman asymptotic conditions are used. Emphasis is therefore placed on the versions of the theorem that satisfy the Feynman asymptotic conditions. In Sect. IV, we discuss a charged scalar field in an external electromagnetic field, prove the appropriate version of Kirchhoff's integral theorem, and demonstrate that in an external electromagnetic field, the superposition schemes are not fundamentally modified.

In nonlinear theories the superposition principle holds in relation to the secondary waves.
In Sect. V, we consider a class of nonlinear scalar field theories.
The physical meaning of Kirchhoff's integral theorem is discussed,
including its connections with the GF method and the superposition principle.
Vectorial generalizations of Kirchhoff's integral theorem for retarded Green's function
are discussed Appendix.
The conclusions section summarizes the discussion.

\section{The Huygens-Fresnel superposition principle and Kirchhoff's integral theorem in the oscillator problem}
\renewcommand{\theequation}{II.\arabic{equation}}
\setcounter{equation}{0}

A free scalar field obeys the Klein-Gordon equation:
\begin{equation}
(\Box +m^{2} )\phi _{0} ^{} (x)=0.
\label{1}
\end{equation}

Of interest are the general features of solutions of the wave equation,
which extend to its nonlinear modifications.
The main consequences of Kirchhoff's theorem and the physical content of the Fresnel-Huygens superposition principle can be explained using the example of the one-dimensional oscillator; thus, we begin by considering the evolution of a one-dimensional harmonic oscillator. This problem can also be regarded as a problem of the evolution of a free scalar field in momentum space.

\subsection{Harmonic oscillator}

We write the equation in the form
\begin{equation}
\left(\frac{d^{2} }{dt^{2} } +m^{2} \right)\phi _{0} (t)=0.
\label{2}
\end{equation}
Here, $m$ is the frequency of the oscillator and $\phi _{0} (t)$ is its coordinate. If $\phi _{0} (t)$ is a spatially homogeneous field in the Klein-Gordon equation, then $m$ is the mass of the particle.

\subsubsection{Complete orthonormal basis functions}

A complete set of solutions to Eq.~(\ref{2}) is formed by the two functions
\begin{equation}
f^{(+)} (t)=\frac{e^{-imt} }{\sqrt{2m} } \, \, \, \, \textrm{and} \, \, \, \, f^{(-)} (t)=\frac{e^{imt} }{\sqrt{2m} } .
\label{3}
\end{equation}
The normalization and completeness conditions are expressed in terms of the Wronskian. If $\varphi $ and $\chi $ are two functions, then their Wronskian is equal to
\begin{equation}
W[\varphi ,\chi ]=\det \left\| \begin{array}{cc} {\varphi } & {\chi } \\ {\dot{\varphi }} & {\dot{\chi }} \end{array}\right\| =\varphi \dot{\chi }-\dot{\varphi }\chi .
\end{equation}
The notation
\[\varphi \stackrel{\leftrightarrow}{\partial }_{t} \chi =W[\varphi ,\chi ]\]
is often used. The normalization and orthogonality of the basis functions are represented as follows:
\begin{equation}
iW[f^{(\pm )*} ,f^{(\pm )} ]=\pm 1 \, \, \, \, \textrm{and} \, \, \, \, W[f^{(\pm )*} ,f^{(\mp )} ]=0. \label{4}
\end{equation}
If the functions for which we compute the Wronskian are solutions of Eq.~(\ref{2}), then the Wronskian is independent of time. Let $\phi_{0} (t)$ be a solution of Eq.~(\ref{2}). We define the following time-independent complex numbers:
\begin{equation}
a=iW[f^{(+)*} ,\phi _{0} ] \, \, \, \, \textrm{and} \, \, \, \, a^{*} =-iW[f^{(-)*} ,\phi _{0} ]. \label{5}
\end{equation}
After quantization, the values $a$ and $a^*$ become annihilation and creation operators.

The completeness condition takes the form
\begin{equation}
\phi _{0} (t)=f^{(+)} (t)iW[f^{(+)*} ,\phi _{0} ]-f^{(-)} (t)iW[f^{(-)*} ,\phi _{0} ]. \label{6}
\end{equation}
This equation also allows for the decomposition of the solution into its positive- and negative-frequency components:
\begin{equation}
\phi _{0}^{} (t)=\phi _{0}^{(+)} (t)+\phi _{0}^{(-)} (t), \label{7}
\end{equation}
where
\begin{equation}
\phi_{0}^{(\pm )} (t)=\pm f^{(\pm )} (t)iW[f^{(\pm )*} ,\phi _{0} ].\end{equation}


Equation (\ref{6}) is valid not only in the linear vector space spanned by the basis functions (\ref{3}), but also for any function evaluated at time $t$. The right-hand side of Eq.~(\ref{6}) for an arbitrary function $\chi(t)$ has the form
\begin{equation*}
\textrm{r.h.s.}
=i\left( f^{(+)}(t)f^{(+)\ast }(t)-f^{(-)}(t)f^{(-)\ast
}(t)\right) \dot{\chi}(t)
-i\left( f^{(+)}(t)\dot{f}^{(+)\ast }(t)-f^{(-)}(t)\dot{f}%
^{(-)\ast }(t)\right) \chi(t).
\end{equation*}
Using the explicit form of $f^{(\pm)}(t)$, one can see that $\textrm{r.h.s.} = \chi(t)$. Although this property appears fortuitous, it is rather fundamental.

Let us consider the Poisson bracket relations
\begin{eqnarray}
\{\phi _{0}(t),\phi _{0}(t)\} &=&0, \label{PB1} \\
\{\phi _{0}(t),\pi _{0}(t)\} &=&1,  \label{PB2}
\end{eqnarray}
where $\pi_{0}(t) = \dot{\phi}_{0}(t)$ is the canonical momentum. A simple calculation using Eq.~(\ref{6}) gives
\begin{eqnarray}
\{\phi _{0}(t^{\prime }),\phi _{0}(t)\}
&=&f^{(+)}(t)i\{\phi _{0}(t^{\prime }),W[f^{(+)\ast },\phi
_{0}]\}-f^{(-)}(t)i\{\phi _{0}(t^{\prime }),W[f^{(-)\ast },\phi _{0}]\} \nonumber \\
&=&f^{(+)}(t)i\{\phi _{0}(t^{\prime }),f^{(+)\ast }(t^{\prime })\pi
_{0}(t^{\prime })-\dot{f}^{(+)\ast }(t^{\prime })\phi _{0}(t^{\prime
})\} \nonumber \\
&-&f^{(-)}(t)i\{\phi _{0}(t^{\prime }),f^{(-)\ast }(t^{\prime })\pi
_{0}(t^{\prime })-\dot{f}^{(-)\ast }(t^{\prime })\phi _{0}(t^{\prime })\} \nonumber \\
&=& i\left(
f^{(+)}(t)f^{(+)\ast}(t^{\prime})-f^{(-)}(t)f^{(-)\ast}(t^{\prime})
\right), \label{bracket1} \\
\{\phi _{0}(t^{\prime }),\pi  _{0}(t)\}
&=& i\left(
\dot{f}^{(+)}(t)f^{(+)\ast }(t^{\prime }) - \dot{f}^{(-)}(t) f^{(-)\ast }(t^{\prime})
\right). \label{bracket2}
\end{eqnarray}
By virtue of Eqs.~(\ref{PB1}) and (\ref{PB2}),
\begin{eqnarray}
f^{(+)}(t)f^{(+)\ast}(t)-f^{(-)}(t)f^{(-)\ast}(t)
 &=&0, \label{Fund1} \\
f^{(+)}(t)\dot{f}^{(+)\ast }(t) - f^{(-)}(t)\dot{f}^{(-)\ast }(t)
 &=& i, \label{Fund2} \\
\dot{f}^{(+)}(t)f^{(+)\ast }(t) - \dot{f}^{(-)}(t)f^{(-)\ast }(t)  &=& -i. \label{Fund3}
\end{eqnarray}

Identity $\textrm{r.h.s.} = \chi(t)$ is, therefore, a consequence of the completeness condition (\ref{6}) for functions $\phi_0(t)$, which are solutions of Eq.~(\ref{2}), and the Poisson bracket relations for the canonical variables.

\subsubsection{The Green's functions}

A Green's function is defined by the equation
\begin{equation}
\left(\frac{d^{2} }{dt^{2} } +m^{2} \right)\Delta_X (t)=-\delta (t). \label{8}
\end{equation}
By performing the Fourier transform in time, we obtain the Green's function in frequency space:
$\Delta_X (\omega )= (\omega ^{2} -m^{2})^{-1}$.
For the inverse Fourier transformation,
\begin{equation}
\Delta_X (t)=\int _{-\infty }^{+\infty }\frac{d\omega }{2\pi }  \, e^{-i\omega t} \frac{1}{\omega ^{2} -m^{2} } ,
\label{10}
\end{equation}
it is necessary to bypass the poles on the real axis that arise for $\omega =\pm m.$ There are four possibilities, which correspond to four Green's functions:
\begin{eqnarray}
\Delta_{F}(t'-t) &=& \int_{-\infty}^{+\infty}\frac{d\omega }{2\pi}
e^{-i\omega (t'-t)} \frac{1}{ \omega^{2} - m^{2} +i0}
\nonumber \\
&=&-i\left( f^{(+)}(t') f^{(+)*}(t)\theta(t'-t) + f^{(-)}(t')f^{(-)*}(t)\theta(-t'+t)\right),
\label{11} \\
\Delta _{F} ^{c} (t'-t)&=&\int _{-\infty }^{+\infty }\frac{d\omega }{2\pi }  \, e^{-i\omega (t'-t)} \frac{1}{\omega ^{2} -m^{2} -i0}  \nonumber \\
&=& i\left(f^{(-)} (t')f^{(-)*} (t)\theta (t'-t)+f^{(+)} (t')f^{(+)*} (t)\theta (-t'+t)\right),
\label{12} \\
\Delta _{\mathrm{ret}} (t'-t)&=&\int _{-\infty }^{+\infty }\frac{d\omega }{2\pi } e^{-i\omega (t'-t)} \frac{1}{\omega ^{2} -m^{2} +i0\mathrm{sgn}(\omega )}  \nonumber \\
&=&-i\left( f^{(+)} (t')f^{(+)*} (t)-f^{(-)} (t')f^{(-)*} (t)\right)\theta (t'-t),
\label{13} \\
\Delta _{\mathrm{adv}} (t'-t)&=&\int _{-\infty }^{+\infty }\frac{d\omega }{2\pi }  e^{-i\omega (t'-t)} \frac{1}{\omega ^{2} -m^{2} -i0\mathrm{sgn}(\omega )}  \nonumber \\
&=& i\left(f^{(+)} (t')f^{(+)*} (t)-f^{(-)} (t')f^{(-)*} (t)\right)\theta (-t'+t).
\label{14}
\end{eqnarray}
Each of these functions satisfies Eq.~(\ref{8}). The difference between any two Green's functions is a solution of the free equation (\ref{2}).

It is instructive to verify by the direct calculation that the representation (\ref{11}) satisfies Eq.~(\ref{8}). With the help of equation
\begin{equation*}
f(x)\delta ^{\prime }(x)= f(0)\delta ^{\prime }(x)-f^{\prime }(0)\delta (x),
\end{equation*}%
one finds
\begin{eqnarray}
\left( \frac{d^{2}}{dt^{\prime 2}} + m^{2} \right) i\Delta_F (t^{\prime }-t) &=&
2 \left( \dot{f}^{(+)}(t^{\prime })f^{(+)\ast }(t) - \dot{f}^{(-)}(t^{\prime })f^{(-)\ast }(t)\right) \delta (t^{\prime }-t) \nonumber \\
&+&\left( f^{(+)}(t^{\prime })f^{(+)\ast}(t)
-f^{(-)}(t^{\prime })f^{(-)\ast}(t)\right) \delta^{\prime }(t^{\prime }-t) \nonumber \\
&=&
\left( \dot{f}^{(+)}(t^{\prime })f^{(+)\ast }(t)
-\dot{f}^{(-)}(t^{\prime })f^{(-)\ast }(t) \right) \delta (t^{\prime }-t) \nonumber \\
&+&
\left( f^{(+)}(t)f^{(+)\ast }(t)
-f^{(-)}(t)f^{(-)\ast }(t)\right) \delta ^{\prime }(t^{\prime }-t).
\label{delta1}
\end{eqnarray}%
Using Eqs.~(\ref{Fund1}) and (\ref{Fund3}), we arrive at Eq.~(\ref{8}).

In terms of quantized variables, the Feynman propagator is defined by
\begin{equation}
i\Delta _{F}(t^{\prime }-t)=\langle 0|T\hat{\phi}_{0}(t^{\prime })\hat{\phi}_{0}(t)|0\rangle.
\label{A7}
\end{equation}
The $T$ product entering this expression
occurs naturally in solutions of the evolution equation
$i\partial_{t} \Psi(t) = \hat{H}(t) \Psi(t)$
of systems with a time-dependent Hamiltonian.
If, at various times, $\hat{H}$ does not commute with itself, namely $[\hat{H}(t'),\hat{H}(t)] \neq 0$, then
the solution $\Psi(t) = U(t,0) \Psi(0)$ is expressed in terms of the time-ordered exponential
$U(t,0) = T \exp (-i\int^{t}_{0} \hat{H}(t^{\prime})dt^{\prime})$.
In perturbation theory $\Delta _{F}(t^{\prime }-t)$ then arises by Wick's theorem,
which explains why $\Delta _{F}(t^{\prime }-t)$ plays a special role in quantum theory.
The definition (\ref{A7}) is consistent with the definition (\ref{11}).

\subsubsection{Superposition principle from Kirchhoff's integral theorem}

Let us compute the Wronskian of the Feynman propagator $\Delta _{F}^{} (t^{\prime} -t)$ and a solution $\phi _{0} (t)$ of Eq.~(\ref{2}). By taking the derivative with respect to $t$ of $W[\Delta _{F} (t^{\prime} -t),\phi _{0} (t)]$ and integrating the result over the interval $(t_{1} ,t_{2} )$, the following equation is obtained for $t_{1} < t^{\prime} < t_{2} $:
\begin{equation}
\phi _{0} (t^{\prime} )=W[\Delta _{F}^{} (t^{\prime} -t_{2} ),\phi _{0} (t_{2} )]-W[\Delta _{F}^{} (t^{\prime} -t_{1} ),\phi _{0} (t_{1} )].
\label{15}
\end{equation}

This relation is the harmonic oscillator analog of Kirchhoff's integral theorem.
Despite the drastic simplification, the fundamental meaning is maintained and is amenable to interpretation. According to Eq.~(\ref{15}), the coordinate $\phi_{0}(t)$ is determined by both the past and the future.
From the past, the Wronskian selects the positive-frequency component of $\phi _{0} (t_{1} )$ and propagates it into the future up to the moment $t=t^{\prime} >t_{1} $. From the future, the Wronskian selects the negative-frequency component of $\phi _{0} (t_{2} )$ and propagates it into the past up to the moment $t=t^{\prime} <t_{2} $. The result is a superposition of the two
\textit{waves}.
Equation (\ref{2}) is commonly regarded as the equation of motion of a particle (oscillator) in the 1-dimensional space. A less obvious interpretation of this equation as an evolution equation of a wave in the 0-dimensional space is also possible. Equation (\ref{15}) underlines the second interpretation.

The analogy with quantum field theory is apparent: particles are identified with positive-frequency solutions of wave equations, and antiparticles are identified with negative-frequency solutions. Particles move \textit{forward in time}, whereas antiparticles move \textit{backward in time}. In accordance with the Huygens-Fresnel superposition principle adapted here
for the Feynman asymptotic conditions,
the wave $\phi _{0} (t^{\prime} )$ is equal to the sum of the negative-frequency component of $\phi _{0} (t_{2} )$, propagating backward in time, and the positive-frequency component of $\phi _{0} (t_{1} )$, propagating forward in time. Equation (\ref{15}) can thus be interpreted both in the spirit of the Huygens-Fresnel superposition principle and in the spirit of the GF method, thereby establishing the close relationship between them.

According to Eq.~(\ref{15}), the coordinate $\phi _{0} (t^{\prime} )$ is determined by its value and its first derivative at the other two time points. Arguing reversely, this suggests that the evolution equation contains time derivatives of no higher than second order.

If $t^{\prime} \notin (t_{1} ,t_{2} )$, then there is a zero on the left-hand side of Eq.~(\ref{15}):
\begin{equation}
0=W[\Delta _{F}(t^{\prime} -t_{2} ),\phi _{0} (t_{2} )]
- W[\Delta _{F}(t^{\prime} -t_{1} ),\phi _{0} (t_{1} )].
\label{16}
\end{equation}

Equations (\ref{15}) and (\ref{16}) remain valid after the replacement $\Delta _{F}$ with any other propagator. For the retarded Green's function, the analog of Eqs.~(\ref{15}) and (\ref{16}) for $t_2 \to +\infty$ reads
\begin{equation}
\phi _{0}^{} (t^{\prime} )\theta (t^{\prime} - t_{1})=-W[\Delta _{\mathrm{ret}}^{} (t^{\prime} -t_{1}),\phi _{0}^{} (t_{1})].
\label{17}
\end{equation}
Here, the positive- and negative-frequency components propagate forward in time, corresponding to the usual formulation of the Huygens-Fresnel superposition principle, so that $\phi_{0}(t)$ is determined by the past only.

\subsubsection{Superposition principle from the completeness condition}

Here, we present a different formulation of the 
superposition principle. To begin, let us find the Wronskian $W$ of $\Delta _{F} (t^{\prime} -t)$ and $\phi _{0} (t)$. The expression (\ref{11}), when substituted into $W$, yields
\begin{eqnarray}
 W[\Delta _{F}(t^{\prime} -t),\phi _{0} (t)] &=&
   - if^{(+)}(t^{\prime} )W[f^{(+)*} (t)\theta(t^{\prime} -t),\phi _{0} (t)]
   -if^{(-)}(t^{\prime} )W[f^{(-)*} (t)\theta(t-t^{\prime} ),\phi _{0}] \nonumber \\
&=&-if^{(+)}(t^{\prime} )\theta(t^{\prime} -t)W[f^{(+)*} ,\phi _{0} ]-if^{(-)} (t^{\prime} )\theta (t-t^{\prime} )W[f^{(-) *} ,\phi _{0} ] \nonumber \\
&& + \Delta(t^{\prime}  - t) \phi _{0} (t)\delta (t^{\prime} -t),
\label{18}
\end{eqnarray}
where
\begin{equation}
i\Delta(t^{\prime}  - t) = f^{(+)}(t^{\prime} )f^{(+) *} (t)-f^{(-)} (t^{\prime} )f^{(-) *} (t).
\label{comm}
\end{equation}
By virtue of Eq.~(\ref{bracket1})
\begin{equation*}
\Delta(t^{\prime }  - t) = \{\phi _{0}(t^{\prime }),\phi _{0}(t)\}.
\end{equation*}
In the transition to the last lines of Eq.~(\ref{18}), the properties of the Wronskian and the definitions of the basis functions (\ref{3}) are used.
According to Eq.~(\ref{Fund1}), the term $\sim \Delta(t^{\prime }  - t) \delta(t^{\prime }  - t)$ vanishes, yielding
\begin{equation}
\phi _{0}^{(+)} (t^{\prime} )\theta (t^{\prime} -t)-\phi _{0}^{(-)} (t^{\prime} )\theta (t-t^{\prime} )=-W[\Delta _{F}^{} (t^{\prime} -t),\phi _{0}^{} (t)].
\label{19}
\end{equation}

Equation (\ref{19}) can be regarded as an equation for
$\Delta _{F}(t^{\prime} -t)$. By taking the time ($t$) derivative of both sides, we obtain Eq.~(\ref{8}). The superposition principle, formalized as in (\ref{19}), thus determines the Green's function up to a solution of the free equation. To obtain a unique Green's function, the asymptotic behavior must be fixed. By taking the differences between both sides of Eq.~(\ref{19}) for $t=t_{2} $ and $t=t_{1}< t_{2}$, we obtain Eq.~(\ref{15}), provided that $t^{\prime} \in (t_{1} ,t_{2} )$. If the inverse condition, $t^{\prime} \notin (t_{1} ,t_{2} )$, holds, then we obtain Eq.~(\ref{16}). Finally, by taking the time ($t^{\prime} $) derivative, we obtain the superposition principle for the canonical momentum $\pi _{0}^{} (t)=\dot{\phi }_{0}^{} (t)$:
\begin{equation}
\pi _{0}^{(+)} (t^{\prime} )\theta (t^{\prime} -t)-\pi _{0}^{(-)} (t^{\prime} )\theta (t-t^{\prime} )=-W[\Delta _{F}^{} (t^{\prime} -t),\pi _{0}^{} (t)].
\label{19momentum}
\end{equation}

The proof of Eq.~(\ref{19}) is not based on Kirchhoff's theorem nor its obvious modification. For the retarded Green's function, the completeness condition does not lead to a new equation (compared with (\ref{17})).
In quantum field theory, the diagram technique is based on the Feynman propagator; thus, what is of interest to us here is the superposition principle formalized as in (\ref{15}), (\ref{16}) and (\ref{19}).

\subsubsection{Path integral}

Kirchhoff's integral theorem can also be used as a starting point for developing path integral method.

To show this, we note a useful relation
\begin{eqnarray}
iW[\Delta _{F}(t_{3}-t_{2}),\Delta _{F}(t_{2}-t_{1})]&=&-\theta
(t_{3}-t_{2})\theta (t_{2}-t_{1})f^{(+)}(t_{3})f^{(+)\ast }(t_{1}) \nonumber \\
&&+\theta(t_{1}-t_{2})\theta (t_{2}-t_{3})f^{(-)}(t_{3})f^{(-)\ast }(t_{1}).
\label{twopro}
\end{eqnarray}%
This relation tells that a wave propagating toward the future continues to propagate forward in time.
A similar property holds for waves propagating backward in time. 
We choose a sequence of the intervals
$(t_{1},t_{2})\subset (t_{3},t_{4})\subset \ldots \subset
(t_{2n-1},t_{2n})$ and consider $t^{\prime} \in
(t_{1},t_{2})$. Eq.~(\ref{15}) being iterated $n$ times gives
\begin{eqnarray}
\phi _{0}(t^{\prime})&=& W[\Delta _{F}(t^{\prime }-t_{2}),W[\Delta
_{F}(t_{2}-t_{4}),W[\ldots ,W[\Delta _{F}(t_{2n}-t_{2n + 2}),\phi_{0}(t_{2n + 2})]\ldots ]]] \label{pi} \\
&+&(-)^{n + 1}W[\Delta _{F}(t^{\prime }-t_{1}),W[\Delta_{F}(t_{1}-t_{3}),W[\ldots ,W[\Delta _{F}(t_{2n-1}-t_{2n+1}),\phi_{0}(t_{2n+1})]\ldots ]]]. \nonumber
\end{eqnarray}
According to this equation, $\phi _{0}(t_{2n+2})$ generates a secondary wave that propagates into the past.
In the neighboring instant of time $t=t_{2n}< t_{2n+2}$ it generates new secondary wave, etc.
The same interpretation is valid for the wave propagating forward in time.
Equation~(\ref{15}) is reproduced with $n=0$ for $t_{-1} = t_{0} = t^{\prime}$.
The mixed terms containing forward and backward propagation do not arise as a consequence of (\ref{twopro}).
In the limit of $n \rightarrow \infty $, $t_2 - t_1 \to 0$ and $(t_{l+3} - t_{l+2}) \to (t_{l+1} - t_{l})$, we arrive at the continuous product \textit{over history}.
Equation (\ref{pi}) can be regarded as a path-integral representation in the space $\mathbb{R}^{1,0}$.

Path integral in the space $\mathbb{R}^{1,3}$ is discussed in Sect. III.E.

\subsection{Harmonic oscillator with a time-dependent frequency }

A field theoretical version of the evolution problem with a time-dependent oscillator frequency, in light of the superposition principle, is discussed in Sect. IV, where proofs are presented. Here, we restrict ourselves to statements of the main assertions.

We consider the equation
\begin{equation}
\left(\frac{d^{2} }{dt^{2} } +m^{2} +\Delta m^{2} (t)\right)\phi (t)=0,
\label{20}
\end{equation}
where $\Delta m^{2} (\pm \infty )=0.$ The perturbation $\Delta m^{2} (t)$ is switched on and off adiabatically. Let $\Delta _{F} (t',t)$ be the Feynman propagator for Eq.~(\ref{20}). The following superposition schemes hold: As a consequence of Kirchhoff's integral theorem,
\begin{eqnarray*}
\phi (t^{\prime} )&=& W[\Delta_{F}(t^{\prime} ,t_{2}),\phi (t_{2} )]
              - W[\Delta_{F}(t^{\prime} ,t_{1}),\phi (t_{1} )]
\;\;\;\;\textrm{for} \;\;\;\; t^{\prime} \in (t_{1} ,t_{2} ),  \\
          0&=&  W[\Delta_{F}(t^{\prime} ,t_{2}),\phi (t_{2} )]
               -W[\Delta_{F}(t^{\prime} ,t_{1}),\phi (t_{1} )] \;\;\;\; \textrm{for} \;\;\;\; t^{\prime} \notin (t_{1} ,t_{2} ),
\end{eqnarray*}
and, as a consequence of the completeness condition,
\begin{equation*}
\phi _{}^{(+)} (t^{\prime} )\theta (t^{\prime} -t)-\phi _{}^{(-)} (t^{\prime} )\theta (t-t^{\prime} )=-W[\Delta _{F}^{} (t^{\prime} ,t),\phi (t)],
\end{equation*}
where $\phi^{(\pm )} (t) \sim f^{(\pm )} (t)$ at $t\to \pm \infty $.  The expansion of $\phi (t)$ into positive- and negative-frequency components $ \phi^{(\pm)} (t)$ has an objective meaning because the evolution equation is linear.

\subsection{Anharmonic oscillator}

Let us consider a more general case. We add to the oscillator potential an arbitrary potential $V(\phi )$. The equation of motion takes the form
\begin{equation}
\left(\frac{d^{2} }{dt^{2} } +m^{2} \right)\phi (t)=-V'(\phi (t)).
\label{21}
\end{equation}

\subsubsection{Superposition principle from Kirchhoff's integral theorem}

Equation (\ref{15}) is modified as follows:
\begin{equation}
\phi (t^{\prime} )=W[\Delta _{F}^{} (t^{\prime} -t_{2} ),\phi (t_{2} )]-W[\Delta _{F}^{} (t^{\prime} -t_{1} ),\phi (t_{1} )]+\int _{t_{1} }^{t_{2} }dt\Delta _{F}^{} (t^{\prime} -t)V'(\phi (t)) .
\label{22}
\end{equation}
The propagator $\Delta _{F}^{} (t)$ is determined from Eq.~(\ref{8}). On the interval $(t_{1} ,t_{2} )$ the sum of the first two terms satisfies the evolution equation of the harmonic oscillator. We denote this sum as
\begin{equation}
\phi _{0} (t^{\prime} )\equiv W[\Delta _{F}^{} (t^{\prime} -t_{2} ),\phi (t_{2} )]-W[\Delta _{F}^{} (t^{\prime} -t_{1} ),\phi (t_{1} )].
\label{23}
\end{equation}
The solution takes the form
\begin{equation}
\phi (t^{\prime} )=\phi _{0} (t^{\prime} )+\int _{t_{1} }^{t_{2} }dt\Delta _{F}^{} (t^{\prime} -t)V'(\phi (t)) .
\label{24}
\end{equation}
Given that the Green's function properties of the harmonic oscillator are known, the solution can be written immediately. If $t^{\prime} \notin (t_{1} ,t_{2} )$, then we obtain
\begin{equation}
0=\phi _{0} (t^{\prime} )+\int _{t_{1} }^{t_{2} }dt\Delta _{F}^{} (t^{\prime} -t)V'(\phi (t)) .
\label{25}
\end{equation}

The last two equations constitute a version of Kirchhoff's integral theorem for the one-dimensional anharmonic oscillator.

Equation (\ref{24}) cannot be interpreted canonically. Although the first term has the standard meaning under the Fresnel superposition scheme, the second term indicates that a component arises among the secondary waves that is generated continuously in time.

According to the Huygens-Fresnel superposition principle, to describe the propagation of a wave, it is sufficient to know its phase and amplitude at a fixed time. However, this is true only in linear theories. In nonlinear theories, the propagation of a wave
is determined by its entire history (for retarded solutions, its prehistory), even if the original wave equation is local. The dependence of the wave observables on the entire history of the wave indicates, in general, the nonlocal nature of its evolution. Only a narrow family of representations that contain an integral over time correspond to local but nonlinear theories.

The derivative of the potential is an additional source of secondary waves (corrections to the coordinate), and the potential depends on the exact coordinate. This means that Eq. (\ref{24}) is self-consistent and that its solution is obvious only in the context of perturbation theory.

In quantum field theory, an equation similar to Eq.~(\ref{24}) serves as the starting point for the development of the diagram technique (see, e.g., \cite{Bjor64}). The equations obtained by replacing the Feynman propagator in Eq.~(\ref{24}) with the retarded and advanced propagators are used to develop the axiomatic scattering theory
(see, e.g., \cite{Bjor65}).

\subsubsection{Positive- and negative-frequency solutions}

In the theory of interacting fields, the decomposition of solutions into positive- and negative-frequency components makes sense only asymptotically for outgoing and incoming states. We assume that the nonlinear interaction is adiabatically switched on at $t \to - \infty$ and adiabatically switched off at $t \to + \infty$. If positive- and negative-frequency components $\phi^{(\pm )} (t)$ are somehow defined, then the subsequent modification of Eq.~(\ref{19}) is obvious:
\begin{equation}
\phi^{(+)}(t^{\prime} )\theta (t^{\prime} -t)-\phi _{}^{(-)} (t^{\prime} )\theta (t-t^{\prime} )=-W[\Delta _{F}^{} (t^{\prime} -t),\phi (t)]
+\int _{t}^{t^{\prime} }d\tau\Delta _{F} (t^{\prime} - \tau)V'(\phi (\tau)) .
\label{26}
\end{equation}

By taking the time ($t$) derivative, after some simple transformations, we obtain $\phi (t)=\phi^{(+)} (t)+\phi^{(-)} (t)$ and Eq.~(\ref{8}). The difference in this equation at two unequal time points leads to Eqs.~(\ref{22}) - (\ref{25}). It might seem, therefore, that Eq. (\ref{26}) is no less general than Eqs.~(\ref{22}) - (\ref{25}). However, we do not have an independent definition of the decomposition into positive- and negative-frequency components. We are forced, therefore, to regard Eq.~(\ref{26}) as a definition of $\phi^{(\pm)} (t)$. According to this equation, $\phi _{}^{(\pm )} (t)\sim f^{(\pm )} (t)$ at $t\to \pm \infty $.

The calculation of the first derivative of Eq.~(\ref{26}) in $t^{\prime} $ leads to the superposition principle for the canonical momentum
\begin{equation}
\pi^{(+)} (t^{\prime} )\theta (t^{\prime} -t)-\pi _{}^{(-)} (t^{\prime} )\theta (t-t^{\prime} )=-W[\Delta _{F}(t^{\prime} -t),\pi (t)]
+\int _{t}^{t^{\prime} }d\tau\Delta _{F}^{} (t^{\prime} -\tau)V''(\phi (\tau))\pi (\tau) .
\end{equation}
This equation is consistent with the evolution equation for $\pi^{(\pm )} (t)=\dot{\phi }^{(\pm )}(t)$.

Obviously, in nonlinear theories, a full generalization of (\ref{19}) does not exist.

A field theoretical version of the anharmonic oscillator problem is discussed in Sect. V.

\subsubsection{Numerical example}

We use a numerical example to demonstrate the application of the superposition scheme (\ref{24}) for the description of radial motion in the Keplerian problem.
After separation of the angular variables, the evolution problem reduces to solving a problem of one-dimensional motion in an effective potential
\[
U=-\frac{\alpha }{r}+\frac{L^{2}}{2\mu r^{2}},
\]%
where $\alpha =GM_{\odot }\mu $, $M_{\odot }$ is the solar mass, $\mu $ is
the mass of a celestial body, and $L$ is the angular momentum.
We add and subtract
from the potential $U$ an oscillator potential
\[
U_{osc}=\frac{1}{2}\mu m^{2}(r-a)^{2}
\]%
and treat $U_{osc}$ as the undisturbed potential.
The perturbation potential is thus $V=U-U_{osc}.$
In order to improve convergence and eliminate the need to determine optimized $U_{osc}$,
the frequency parameter $m$ is chosen in agreement with the exact solution (see, e.g., \cite{Arno89}):
$m=2\pi/T$, where $T=2\pi \mu ab/L$ is the orbital period, $a = (r_{\min } + r_{\max })/2$ and $b = \sqrt{pa}$ are the major and minor semi-axes of the ellipse
and $L = \sqrt{p \alpha \mu}$; the variable $r$ lies in the interval $(r_{\min },r_{\max})$, where $r_{\min } = p/(1 + e)$, $r_{\max } = p/(1 - e)$,
$p$ is the semi-latus rectum, and $e$ is the
eccentricity.

As a zeroth-order approximation for $\phi(t) \equiv r(t) - a$, we choose a free solution
\begin{equation}
\phi^{[0]}(t) = C_{+}^{[0]}f^{(+)}(t)+C_{-}^{[0]}f^{(-)}(t)
\end{equation}
with unknown coefficients $C_{\pm}^{[0]}$ and $f^{(\pm)}(t)$ defined by Eq.~(\ref{3}).
The motion begins at perihelion $\phi^{[0]}(0)=r_{\min } - a$,
with the vanishing velocity $\dot{\phi}^{[0]}(0)=0$. These conditions allow $C_{\pm}^{[0]}$ to be fixed.

Given the $l$th-order approximation, $r^{[l]} (t) = a + \phi^{[l]}(t)$ can be substituted in place of the argument of $V^{\prime }$
in Eq.~(\ref{24}) to produce the next-order iteration
\begin{equation}
\phi^{[l+1]}(t) = C_{+}^{[l+1]}f^{(+)}(t)+C_{-}^{[l+1]}f^{(-)}(t)+\int_{t_{1}}^{t_{2}}d%
\tau \Delta _{F}(t-\tau )V^{\prime }(a + \phi^{[l]}(\tau )),  \label{25 bis}
\end{equation}%
where $\Delta _{F}(t)$ is defined by (\ref{11}).
The interval $(t_{1},t_{2})$ covers an interval within which we seek
the solution. $C_{\pm}^{[l+1]}$ are fixed
by the conditions $\phi^{[l+1]}(0)=r_{\min } - a$ and $\dot{\phi}^{[l+1]}(0)=0$.


\begin{table}[h]
\caption{Expansion coefficients of free solutions in the unperturbed potential
for the first two iterations and for the exact solution ($l=\infty$).}
\label{tab:2}
\centering
\vspace{2pt}
\begin{tabular}{|c|l|l|}
\hline\hline
 $l$       & $~~~~~~~~~~~~~C^{[l]}_{+}$ & $~~~~~~~~~~~~~C^{[l]}_{-}$ \\
\hline\hline
$0$        & $-0.142872$           & $-0.142872$               \\
$1$        & $-0.155969 - i0.040544$ & $-0.155322 - i0.068246$ \\
$\infty$   & $-0.151619 - i0.033743$ & $-0.151875 - i0.033990$ \\
\hline\hline
\end{tabular}%
\end{table}


The numerical convergence of the recursion is a subtle issue that should be studied separately.
Assuming the convergence of the approximate sequence, we should obtain an identity when using $r(t)$ to evaluate the integral in Eq.~(\ref{24}):
\begin{equation}
\phi^{[\infty]}(t) = C_{+}^{[\infty]}f^{(+)}(t)+C_{-}^{[\infty]}f^{(-)}(t)+\int_{t_{1}}^{t_{2}}d%
\tau \Delta _{F}(t-\tau )V^{\prime }(a + \phi(\tau )).
\label{N2}
\end{equation}%

The exact solutions are parameterized in terms of the eccentric anomaly $E$:
$r=a(1-e\cos E )$ and $t=\sqrt{ma^{3}/\alpha }(E -e\sin E )$, where $t$ is time.
For our numerical estimates, we choose $\alpha = \mu = p = 1$ and $e=0.2$.
The values $t_{1}$ and $t_{2}$ are taken arbitrarily; they correspond to $E_{1} = -1$ and $E_{2} = 7.2$.
The coefficients $C_{\pm}^{[l]}$ for $l=0,1,\infty$ found as described above are presented in Table \ref{tab:2}.
Table \ref{tab:1} shows $r^{[0]}$, $r^{[1]}$ and $r^{[\infty]}$ for seven values of $E \in [ 0,2\pi ].$
The inclusion of the secondary waves generated by the nonlinear source $V^{\prime}$ reduces the standard deviation
$\chi^2 = \sum (r^{[l]} - r)^2$ from $0.0038$ to $0.0015$, whereas $r^{[\infty]}$ coincides with $r$.

Equation~(\ref{24}) can also be derived directly,
under the assumption of $t \in (t_1,t_2)$,
by using the GF method,
whereas Eqs.~(\ref{23}) and (\ref{25}) are specific consequences of Kirchhoff's integral theorem.
We verified that the free term in Eq.~(\ref{N2}) fulfills, numerically, Eq.~(\ref{23})
and checked Eq.~(\ref{25}) for a sample set of time points $t \notin (t_1,t_2)$ as well.

\vspace{4pt}

Summarizing, the idea of Kirchhoff’s integral theorem was explained in this section with a one-dimensional toy model (a harmonic oscillator).
Such a pedagogical approach illustrates formalism while the attempt to draw a physical analogy with well-known phenomena
leads to the seemingly paradoxical observation:
no waves in the $\mathbb{R}^{1,0}$ space, but the superposition principle is there,
and even the problem of celestial mechanics was solved using Kirchhoff’s integral theorem in a technically consistent manner.
A parallelism between classical mechanics and geometrical optics
was regarded as purely formal until the advent of quantum mechanics.
The possibility of solving the problems of classical mechanics using the methods of wave optics seems to be a surprising circumstance.


\begin{table}[t]
\caption{First two iterations $r^{[l]}$ for the approximate solution of the radial equation of motion
as compared to the exact solution $r=r^{[\infty]}$ for seven values of $E \in [0,2\pi]$.
}
\label{tab:1}
\centering
\vspace{2pt}
\begin{tabular}{|c|c|c|c|}
\hline\hline
$E$    & $r^{[0]}$ & $r^{[1]}$ &  $r^{[\infty]}$  \\
\hline\hline
$0$        & 0.8333 & 0.8333 & 0.8333  \\
$\pi/3$    & 0.9079 & 0.9410 & 0.9375 \\
$2\pi/3$   & 1.1131 & 1.1619 & 1.1458 \\
$\pi$      & 1.2500 & 1.2500 & 1.2500 \\
$4\pi/3$   & 1.1132 & 1.1232 & 1.1458 \\
$5\pi/3$   & 0.9079 & 0.9094 & 0.9375 \\
$2\pi$     & 0.8333 & 0.8333 & 0.8333 \\
\hline\hline
\end{tabular}%
\end{table}


\section{ Kirchhoff's integral theorem for a free scalar field}
\renewcommand{\theequation}{III.\arabic{equation}}
\setcounter{equation}{0}

\subsection{Complete orthonormal basis functions}

A complete set of solutions to the Klein-Gordon equation is formed by the functions
\begin{equation*}
f_{\mathbf k}^{(+)}(x) =
\frac{e^{-ikx}} {\sqrt{2\omega_{\mathbf k}}} \;\;\;\;
\mathrm{and} \;\;\;\;
f_{\mathbf k}^{(-)}(x) =
\frac{e^{ ikx}} {\sqrt{2\omega_{\mathbf k}}},
\end{equation*}
where $k=(\omega_{\mathbf k},{\mathbf k})$, $\omega _{\mathbf k} = \sqrt{{\mathbf k}^{2} + m^{2} }$, $x=(t,{\mathbf x}) \in \mathbb{R}^{1,3}$,
and $kx = \omega_{\mathbf k}t - {\mathbf k}{\mathbf x}$. These functions correspond to the positive- and negative-frequency solutions in the oscillator problem. The orthonormality conditions are
\begin{eqnarray}
i\int d{\mathbf x} W[f_{{\mathbf k'}}^{(\pm )*} (x),f_{{\mathbf k}}^{(\pm )} (x)]&=&\pm (2\pi )^{3} \delta ({\mathbf k'}-{\mathbf k}), \nonumber \\
\int d{\mathbf x} W[f_{{\mathbf k'}}^{(\mp )*} (x),f_{{\mathbf k}}^{(\pm )} (x)]&=&0.
\label{27}
\end{eqnarray}

For any function $\phi _{0} (x)$ that is a solution of the Klein-Gordon equation,
\begin{equation}
\phi _{0} (x)=\int \frac{d{\mathbf k}}{(2\pi )^{3} }  \left(f_{{\mathbf k}}^{(+)} (x)i\int d{\mathbf y} W[f_{{\mathbf k}}^{(+)*} (y),\phi _{0} (y)]
-f_{{\mathbf k}}^{(-)} (x)i\int d{\mathbf y} W[f_{{\mathbf k}}^{(-)*} (y),\phi _{0} (y)]\right).
\label{28}
\end{equation}
After the second quantization, the time-independent quantities
\begin{equation*}
a({\mathbf k}) = i\int d{\mathbf y}
W[f_{\mathbf k}^{(+)*}(y),\phi_{0}(y)]\;\;\;\; \mathrm{and}
\;\;\;\; a^{*}({\mathbf k})=-i\int d{\mathbf y} W[f_{\mathbf k}^{(-)*}(y),\phi_{0}(y)]
\end{equation*}
become annihilation and creation operators.

The first and the second terms in Eq.~(\ref{28}) are identified with the positive- and negative-frequency components of $\phi_{0} (x)$. According to the completeness condition (\ref{28}), the solutions of the free equation thereby split into the sum
\begin{equation*}
\phi_{0} (x)=\phi _{0} ^{(+)} (x)+\phi _{0} ^{(-)} (x).
\end{equation*}
This decomposition is analogous to the decomposition of Eq.~(\ref{7}). The orthonormality conditions (\ref{27}) and the completeness condition (\ref{28}) are the generalized equivalents to Eqs. (\ref{4}) and (\ref{6}), respectively, for the oscillator problem.

Using the analogy with equations (\ref{PB1}) - (\ref{Fund3}) and the Poisson bracket relations
\begin{eqnarray}
\{\phi _{0}(x),\phi _{0}(y)\}|_{x^0 = y^0} &=&0, \label{PB3} \\
\{\phi _{0}(x),\pi  _{0}(y)\}|_{x^0 = y^0} &=&
\delta(\mathbf{x} - \mathbf{y}), \label{PB4}
\end{eqnarray}
one can prove that
\begin{eqnarray}
\int \frac{d\mathbf{k}}{(2\pi)^3} \left(
f^{(+)}_{\mathbf{k}}(x)f^{(+)\ast}_{\mathbf{k}}(y) - f^{(-)}_{\mathbf{k}}(x)f^{(-)\ast}_{\mathbf{k}}(y) \right)|_{x^0 = y^0}
 &=&0, \label{Fund4} \\
\int \frac{d\mathbf{k}}{(2\pi)^3} \left(
f^{(+)}_{\mathbf{k}}(x)\dot{f}^{(+)\ast }_{\mathbf{k}}(y) - f^{(-)}_{\mathbf{k}}(x)\dot{f}^{(-)\ast }_{\mathbf{k}}(y)
\right)|_{x^0 = y^0} &=& i\delta(\mathbf{x} - \mathbf{y}), \label{Fund5} \\
\int \frac{d\mathbf{k}}{(2\pi)^3} \left(
\dot{f}^{(+)}_{\mathbf{k}}(x)f^{(+)\ast }_{\mathbf{k}}(y) - \dot{f}^{(-)}_{\mathbf{k}}(x)f^{(-)\ast }_{\mathbf{k}}(y)
\right)|_{x^0 = y^0} &=& -i\delta(\mathbf{x} - \mathbf{y}). \label{Fund6}
\end{eqnarray}
Equations (\ref{Fund4}) and (\ref{Fund5}) can be used to show that the completeness condition (\ref{28}) holds for arbitrary functions at $x^0 = y^0$.

\subsection{Feynman propagator}

The equation for the Feynman propagator is
\begin{equation}
(\Box +m^{2} )\Delta_F (x)=-\delta ^{4} (x).
\label{29}
\end{equation}
It is easiest to find the solution in four-momentum space and then apply the Fourier transform to convert it into coordinate space. Here, as in the oscillator problem, we must shift the contour of the integral over $k^0$ from the real axis in the vicinity of
$k^0 = \pm \omega_{\mathbf k}$. The four possible ways to do so correspond to four Green's functions.

The Feynman propagator can be written as follows:
\begin{eqnarray}
\Delta _{F}(x-y)&=&\int \frac{d^{4} k}{(2\pi )^{4} }  \frac{e^{-ik(x-y)} }{k^{2} -m^{2} +i0} \nonumber \\
&=&-i\int \frac{d{\mathbf k}}{(2\pi )^{3} }  \left( f_{{\mathbf k}}^{(+)} (x)f_{{\mathbf k}}^{(+)*} (y)\theta (x^{0} -y^{0} )+f_{{\mathbf k}}^{(-)} (x)f_{{\mathbf k}}^{(-)*} (y)\theta (- x^{0} + y^{0} ) \right).
\label{Fpro}
\end{eqnarray}
In comparison with Eq.~(\ref{11}), the phase space integral is added here. After the replacement $f^{(\pm)}(t) \to f_{\mathbf k}^{(\pm)}(x)$ and the integration over the phase space in Eqs.~(\ref{12}), (\ref{13}), and (\ref{14}), the form of the other propagators is restored.
Using the analogy with Eq.~(\ref{delta1}) and Eqs.~(\ref{Fund4}) and (\ref{Fund6}), one can verify that the propagator (\ref{Fpro}) satisfies Eq.~(\ref{29}).

\subsection{Superposition principle from Kirchhoff's integral theorem}

\subsubsection{General form of the superposition principle}

We start from the identity
\begin{equation}
\phi _{0} (\xi )\delta ^{4} (\xi -x)=\Delta _{F} (x-\xi )\left((\Box _{\xi } +m^{2} )\phi _{0} (\xi )\right)-\left((\Box _{\xi } +m^{2} )\Delta _{F} (x-\xi )\right)\phi _{0} (\xi ).
\label{30}
\end{equation}
The right-hand side can be written in divergence form as follows:
\begin{equation}
\phi _{0} (\xi )\delta ^{4} (\xi -x)=\frac{\partial }{\partial \xi _{\mu } } \left(\Delta _{F} (x-\xi )\frac{\stackrel{\leftrightarrow}{\partial }}{\partial \xi ^{\mu } } \phi _{0} (\xi )\right).
\label{31}
\end{equation}
By taking the integral over a four-dimensional region $\Omega $ and transforming the right-hand side into a surface integral, the equation
\begin{equation}
\phi _{0} (x)\theta (x\in \Omega )=\int _{\partial \Omega }^{}dS_{\xi }^{\mu }  \left(\Delta _{F} (x-\xi )\frac{\stackrel{\leftrightarrow}{\partial }}{\partial \xi ^{\mu } } \phi _{0} (\xi )\right),
\label{32}
\end{equation}
is obtained, where $\theta (x\in \Omega )$ is the indicator function of $\Omega $:
\begin{equation*}
\theta (x\in \Omega )=\left\{\begin{array}{cc} {1,} & {x\in \Omega \, ,} \\ {0,} & {x\notin \Omega \, .} \end{array}\right.
\end{equation*}
By choosing for the surface $\partial \Omega $ a hyperplane $\xi ^{0} =y^{0} $ in the past, i.e., three-dimensional space at a time $\xi ^{0} = y^{0} <x^{0} $, and a three-dimensional space $\xi ^{0} =z^{0} $ at a time
$\xi ^{0} = z^{0} > x^{0} $ in the future, and then combining these spaces at infinity, where the integral vanishes, we arrive at
\begin{equation}
\phi _{0} (x) =\int d{\mathbf z}W[\Delta _{F} (x-z),\phi _{0} (z)] -\int d{\mathbf y}W[\Delta _{F} (x-y),\phi _{0} (y)] .
\label{33}
\end{equation}
If $x\notin \Omega $, we obtain
\begin{equation}
0 =\int d{\mathbf z}W[\Delta _{F} (x-z),\phi _{0} (z)] -\int d{\mathbf y}W[\Delta _{F} (x-y),\phi _{0} (y)] .
\label{34}
\end{equation}

Equation (\ref{33}) states that $\phi _{0} (x)$ is determined by its past and future. Equation (\ref{34}) suggests that the interference of secondary waves outside the interval $(y^{0} ,z^{0} )$ is strictly destructive.

Equation (\ref{32}) and its consequences (\ref{33}) and (\ref{34}) constitute a version of Kirchhoff's theorem in the most general form;
these equations hold for any choice of propagator.

\subsubsection{Monochromatic field }

The Fourier transform simplifies the superposition scheme of secondary waves. We restrict ourselves to the case of monochromatic, spatially inhomogeneous waves. Consider the following Fourier transforms in time of the scalar field and the Green's function:
\begin{equation}
\phi _{0} (\omega ,{\mathbf x})=\int _{-\infty }^{+\infty }dte^{i\omega t}  \phi _{0} (t,{\mathbf x}),\, \, \, \, \Delta _{F} (\omega ,{\mathbf x})=\int _{-\infty }^{+\infty }dte^{i\omega t}  \Delta _{F} (t,{\mathbf x}).
\label{36}
\end{equation}
They satisfy the equations
\begin{equation*}
(\Delta +{\mathbf k}^{2} )\phi _{0} (\omega ,{\mathbf x})=0,\, \, \, \, (\Delta +{\mathbf k}^{2} )\Delta _{F} (\omega ,{\mathbf x})=\delta ({\mathbf x}),
\end{equation*}
where ${\mathbf k}^{2} =\omega ^{2} -m^{2} $. The right-hand side of the identity
\begin{eqnarray}
\phi _{0} (\omega ,{\boldsymbol\xi })\delta ({\boldsymbol \xi}-{\mathbf x}) &=&
-\Delta _{F} (\omega ,{\mathbf x}-{\boldsymbol\xi})\left((\Delta _{{\xi}} +{\mathbf k}^{2} )\phi _{0} (\omega ,{\boldsymbol\xi})\right) \nonumber \\
&& +\left((\Delta _{{\xi}} +{\mathbf k}^{2} )\Delta _{F} (\omega ,{\mathbf x}-{\boldsymbol\xi})\right)\phi _{0} (\omega ,{\boldsymbol\xi}).
\label{G2I}
\end{eqnarray}
can be represented as the divergence
\begin{equation*}
\phi _{0} (\omega ,{\boldsymbol\xi })\delta ({\boldsymbol\xi }-{\mathbf x})=-\frac{\partial }{\partial \xi ^{\alpha } } \left(\Delta _{F} (\omega ,{\mathbf x}-{\boldsymbol\xi })\frac{\stackrel{\leftrightarrow}{\partial }}{\partial \xi ^{\alpha } } \phi _{0} (\omega ,{\boldsymbol\xi })\right).
\end{equation*}
Integrating over the region $\Omega _{3} $, we obtain von Helmholtz's theorem for the monochromatic field: \cite{Helm1860}
\begin{equation}
\phi _{0} (\omega ,{\mathbf x})\theta ({\mathbf x}\in \Omega _{3} )=-\int _{\partial \Omega _{3} }dS^{\alpha }_{\xi }  \Delta _{F} (\omega ,{\mathbf x}-{\boldsymbol\xi })\frac{\stackrel{\leftrightarrow}{\partial }}{\partial \xi ^{\alpha } } \phi _{0} (\omega ,{\boldsymbol\xi }),
\label{37}
\end{equation}
which is a particular case of the third Green's identity \cite{GGreen1828} and a precursor of Kirchhoff's integral theorem.
The integration is performed over the surface $\partial \Omega _{3} $, which is the boundary of $\Omega _{3} $.
The equation shows that the field at the point ${\mathbf x}$ is determined by its values on any surrounding surface. This surface is not required to be the wave surface. If the point ${\mathbf x}$ lies outside the closed surface, then the integral vanishes. Regardless of the specific form of $\Delta _{F} (\omega ,{\mathbf x})$, we can conclude from the form of the equation alone that if the field $\phi _{0} (\omega ,{\mathbf x})$ satisfies a differential equation, then this equation contains derivatives over the spatial coordinates that are no higher than second order.

Equation (\ref{37}) is used to describe the diffraction phenomena of light \cite{Land71,Born99}.

In the monochromatic, spatially inhomogeneous case, the integration is over the surface rather than over the volume, as in Eq.~(\ref{33}). However, because we are discussing the calculation of the Fourier transform in time, an implicit time integration enters the problem.

\subsubsection{Massless field }

For massless particles the interference scheme for secondary waves simplifies. Let us apply the inverse Fourier transform in Eq.~(\ref{37}):
\begin{equation}
\phi _{0} (t,{\mathbf x})\theta ({\mathbf x}\in \Omega _{3} )=-\int _{\partial \Omega _{3} }dS_{{\xi }}^{\alpha }  \int _{-\infty }^{+\infty }dt' \Delta _{F} (t-t',{\mathbf x}-{\boldsymbol\xi })\frac{\stackrel{\leftrightarrow}{\partial }}{\partial \xi ^{\alpha } } \phi _{0} (t',{\boldsymbol\xi }).
\label{38}
\end{equation}
This equation follows from Eq.~(\ref{32}) if we select for $\Omega $ an infinite cylinder whose spatial section $\Omega _{3} $ is covered by the surface of integration $\partial \Omega _{3} $ and the axis is parallel to the time axis.

As is well known, the propagator $\Delta _{F} (t,{\mathbf x})$ does not vanish outside of the light cone $t^{2} -{\mathbf x}^{2} <0$. This property does not generally violate causality, as $\Delta _{F} (t,{\mathbf x})$ also describes the propagation of the wave surfaces at which the phase remains constant. In the relativistic theory,
the phase velocity $v_{p} \equiv \omega _{{\mathbf k}} /\left|{\mathbf k}\right|\ge 1$ is greater than the speed of light; however, it is the group velocity $v_{g} \equiv \partial \omega _{{\mathbf k}} /\partial \left|{\mathbf k}\right| $$=\left|{\mathbf k}\right|/\omega _{{\mathbf k}} \le 1$ with which the propagation of signals is associated.

In the limit of zero mass, the propagator $\Delta _{F} (t,{\mathbf x})$ takes the following form (see \cite{Bjor65}, Appendix B or Eq.~(29) in Ref. \cite{Zavi79} in the massless limit):
\begin{equation}
\Delta _{F} (t,{\mathbf x})=\frac{i}{4\pi ^{2} } \frac{1}{t^{2} -|{\mathbf x}|^{2} -i0}.
\label{39}
\end{equation}
Substituting (\ref{39}) into (\ref{38}) and taking into account that
\begin{equation*}
\int _{-\infty }^{+\infty }dt' \Delta _{F} (t-t',{\mathbf x})e^{\mp i\omega _{{\mathbf k}} t'} = \frac{1}{4\pi |{\mathbf x}| }
e^{\mp i\omega_{{\mathbf k}} (t \mp |{\mathbf x}|) },
\end{equation*}
we obtain
\begin{eqnarray}
\phi _{0} (t,{\mathbf x})\theta ({\mathbf x}\in \Omega _{3} ) =
\frac{1}{4\pi } \int _{\partial \Omega _{3} }dS_{{\boldsymbol\xi }}^{\alpha }  \left[
- \frac{1}{\rho } \frac{\partial }{\partial \xi ^{\alpha } }
\right.
\left(\phi _{0}^{(+)} (t-\rho /c,{\boldsymbol\xi })+\phi _{0}^{(-)} (t+\rho/c ,{\boldsymbol\xi }) \right)&& \nonumber \\
+\left(\frac{\partial }{\partial \xi ^{\alpha } } \frac{1}{\rho } \right)\left(\phi _{0}^{(+)} (t-\rho /c ,{\boldsymbol\xi })+\phi _{0}^{(-)} (t+\rho /c,{\boldsymbol\xi }) \right)&& \nonumber \\
- \left. \frac{1}{\rho } \frac{\partial \rho }{\partial \xi ^{\alpha } }  \frac{\partial }{c \partial t} \left(\phi _{0}^{(+)} (t-\rho /c,{\boldsymbol\xi })-\phi _{0}^{(-)} (t+\rho /c,{\boldsymbol\xi }) \right)
\right],&&
\label{40}
\end{eqnarray}
where $\rho =\left|{\boldsymbol\xi }-{\mathbf x}\right|$ and in the first term the differentiation with respect to $\xi ^{\alpha } $ does not apply to $\rho $.
The dependence on the speed of light $c$ is here made explicit.

Equation (\ref{40}) represents a general form of Kirchhoff's integral theorem for the Feynman asymptotic conditions. The function is determined by its values on the selected arbitrary closed surface, taking into account the delay of the positive-frequency component and the advancement of the negative-frequency component. This representation is possible because massless particles travel at the speed of light, regardless of their momentum.
\footnote{In Euclidean space of dimension $n \geq 3$ Green's function has the form $\Delta(x) \sim 1/(x^2)^{(n-2)/2}$. Performing a Wick rotation, we find that the Green's function as an analytic function of the variable $ t=x^0$ has two isolated poles in the spaces of even dimension and two root branching points in the spaces of odd dimension. This means that in the massless case the Green's function is effectively localized on the light cone in the spaces of even dimension only. Here an analogue of the representation (\ref{40}) holds. In the spaces of odd dimension, the superposition scheme involves the integration over all spatial coordinates. This property of the Green's function suggests that the requirement of equal phase and group velocities and the speed of light is a necessary but not sufficient condition for the representation of superposition scheme in the form of a surface integral.}
By contrast, the speed of a massive particle depends on its momentum; therefore, the more general representation (\ref{38}) includes the integral over time delay and advance. Kirchhoff's theorem is a precise mathematical formulation of the Huygens-Fresnel superposition principle. A special feature of the Feynman asymptotic conditions is that the negative-frequency components are determined by the future.
An analogue of Eq.~(\ref{40}) for the retarded solutions is the original version of Kirchhoff's integral theorem. It is
briefly outlined in Appendix and discussed in detail in Ref.~\cite{Born99}

\subsection{Superposition principle from the completeness condition}

As a formalization of the superposition principle for the Feynman asymptotic conditions, by analogy with Eq.~(\ref{19}), we can consider
\begin{equation}
 \phi _{0} ^{(+)} (x)\theta ( x^{0} - y^{0} )
-\phi _{0} ^{(-)} (x)\theta (-x^{0} + y^{0} )
=-\int d{\mathbf y}W[\Delta _{F} (x-y),\phi _{0} (y)] .
\label{35}
\end{equation}
The physical content of this equation is quite traditional: At the moment $y^{0} $, the wave is a source of secondary waves, and the propagation from point $y$ to point $x$ is described by $\Delta _{F} (x-y)$. To construct the positive-frequency waves, the past $y^{0} <x^{0} $ must be known, and to construct the negative-frequency waves, the future $x^{0} <y^{0} $ must be known. This property is reflected in the presence of the theta functions on the left-hand side of the equation.

The proof of Eq.~(\ref{35}) is similar to the proof of Eq.~(\ref{19}). It is not based on Kirchhoff's theorem but instead relies on the completeness condition (\ref{28}) and the expansion of the Feynman propagator into plane waves. Given that Eq.~(\ref{35}) is postulated, the Green's function is uniquely determined. Indeed, let us take the derivative over $y^{0} $ on both sides of the equation. After the transformation of the integrand, we obtain Eq.~(\ref{29}); it must then be supplemented by asymptotic conditions.

We take the difference (\ref{35}) for two instants of time, $z^{0}$ and $y^{0} $,  such that $y^{0} < x^{0} < z^{0} $. The result is
Eq.~(\ref{33}). If $x^{0} \notin (y^{0} ,z^{0} )$, then we obtain (\ref{34}).

According to equations (\ref{35}) and (\ref{33}), the field is determined by its values and first derivatives at two time points. This property indicates the local nature of the evolution equation. Arguing backward, since the initial conditions required to determine the field are the field values and the first derivatives, the evolution equation may contain time derivatives of no higher than second order.
Additionally, in Eq.~(\ref{32}), a hypersurface $\partial \Omega $ in the form of an infinite cylinder with its axis parallel to the time axis, can be chosen. In such a case, Kirchhoff's theorem would assert that the wave is determined by its values and gradients on a two-dimensional surface at all times. This version of the theorem indicates the local nature of the evolution equation in the spatial coordinates. The corresponding differential equation may contain derivatives of the spatial coordinates of no higher than second order.

\subsection{Path integral}


The path integral representation is a consequence of Eq.~(\ref{32}).
We choose a set of four-dimensional regions $\Omega _{1}\subset \Omega _{2}\subset \ldots \subset \Omega _{n} \subset \mathbb{R}^{1,3}$.
By iterating Eq.~(\ref{32}), we obtain
\begin{eqnarray}
\phi _{0}(x)\theta (x \in \Omega _{1}) &=& \int_{\partial \Omega _{1}}dS_{\xi
_{1}}^{\mu _{1}}\int_{\partial \Omega _{1}}dS_{\xi _{2}}^{\mu _{2}}\ldots
\int_{\partial \Omega _{n}}dS_{\xi _{n}}^{\mu _{n}}  \label{pir13} \\
&\times& \Delta _{F}(x-\xi _{1})\frac{\overset{\leftrightarrow }{\partial }}{\partial \xi _{1}^{\mu _{1}}}
 \Delta _{F}(\xi _{1}-\xi
_{2})\frac{\overset{\leftrightarrow }{\partial }}{\partial \xi _{2}^{\mu
_{2}}} \ldots  \Delta _{F}(\xi _{n-1}-\xi _{n})\frac{\overset{%
\leftrightarrow }{\partial }}{\partial \xi _{n}^{\mu _{n}}}\phi _{0}(\xi
_{n}). \nonumber
\end{eqnarray}

There exists considerable freedom in choosing $\Omega_{i}$. A similar freedom exists in the factorization of
unitary evolution operator $U(t_2,t_1)$ in quantum mechanics, 
where the equation $U(t_2,t_1) = U (t_2,t) U (t,t_1)$ holds for any instant of time $t \in (t_1,t_2)$.
While the evolution operator is factorized in time, the integration in the path integral goes over the coordinates in three-dimensional space.
Such a representation easily follows from Eq.~(\ref{pir13}).
Indeed, choosing $\Omega_{i}$ to be cylinders with infinite radii and axes parallel to the time axis, we arrive at a representation of this kind.
The broken lines connecting the points $x$ and $\xi_{n} \in \partial \Omega_{n}$ through $\xi_{i} \in \partial \Omega_{i}$ ($i=1,\ldots,n-1$)
form in the continuum limit the class of paths over which the continual integral is defined.
The comparison of Eqs.~(\ref{32}) and (\ref{pir13}) also yields, in the limit of $n \to \infty$, an integral representation
for the Green's function in the form of a continual integral.


\section{Charged scalar field in an external electromagnetic field}
\renewcommand{\theequation}{IV.\arabic{equation}}
\setcounter{equation}{0}

Equations (\ref{32}) and (\ref{35}) and their particular cases were obtained for a free field. The following question arises: which relations can be generalized in the presence of an external field? We restrict ourselves to scalar electrodynamics.

\subsection{Complete orthonormal basis functions}

Substituting the normal derivatives with respect to the space-time coordinates in the Klein-Gordon equation with gauge covariant derivatives,
\begin{equation}
\partial _{\mu } \to D_{\mu } =\partial _{\mu } +ieA_{\mu } \label{41}
\end{equation}
yields the evolution equation for a complex scalar field in an external electromagnetic field,
\begin{equation}
(D_{\mu } D^{\mu } +m^{2} )\phi (x)=0.
\label{42}
\end{equation}

The external field is adiabatically switched on at $t\to -\infty $ and off at $t\to +\infty $. The set of positive- and negative-frequency asymptotic solutions $f_{{\mathbf k}}^{(\pm )} (x)$ is complete and orthonormal. The second-order Eq.~(\ref{43}) has a set of independent solutions $F_{{\mathbf k}}^{(\pm )} (x)$. The asymptotic conditions can be taken as
\begin{equation*}
F_{{\mathbf k}}^{(\pm )} (x)\to f_{{\mathbf k}}^{(\pm )} (x)\equiv \frac{e^{\mp ikx} }{\sqrt{2\omega_{ \mathbf{k} } } } \, \, \, \, {\mathrm {for}}\, \, \, \, t \to -\infty .
\end{equation*}
All other solutions of Eq.~(\ref{42}) are expressed as linear superpositions of the basis functions $F_{{\mathbf k}}^{(\pm )} (x)$.

It would be natural to use the prescription (\ref{41}) for extending the Huygens-Fresnel superposition principle. It can be assumed that in an external electromagnetic field, the suitable generalization of the Wronskian is given by
\begin{equation*}
W_{A} [\varphi ^{*} ,\chi ]\equiv \varphi ^{*} (\stackrel{\leftrightarrow}{\partial }_{t} +2ieA_{0} )\chi =\varphi ^{*} (D_{t} \chi )-(D_{t} \varphi )^{*} \chi .
\end{equation*}
We note a useful property:
\begin{eqnarray}
\partial _{t} W_{A} [\varphi ^{*} ,\chi ]
&=&
\partial _{t} (\varphi ^{*} (D_{t} \chi )-(D_{t}^{} \varphi )^{*}
\chi ) \nonumber \\
&=& \varphi^{*} (D_{t} D_{t} \chi )-(D_{t}^{} D_{t}^{} \varphi )^{*} \chi .
\label{43}
\end{eqnarray}

It is not difficult to show that if $\varphi $ and $\chi $ are two solutions of Eq.~(\ref{42}), then the following condition holds:
\begin{equation*}
\partial_{t} \int d \mathbf{x} W_{A} [\varphi ^{*} ,\chi ]=0.
\end{equation*}
This condition allows us to calculate the normalization integral by sending the time variable to negative infinity, where solutions are represented as plane waves. The orthonormality conditions thus take the form
\begin{eqnarray*}
i\int  d {\mathbf x} W_{A} [F_{{\mathbf k}'}^{(\pm )*} (x),F_{{\mathbf k}}^{(\pm )} (x)]&=&\pm (2\pi )^{3} \delta ({\mathbf k}'-{\mathbf k}), \\
\int  d {\mathbf x} W_{A} [F_{{\mathbf k}'}^{(\pm )*} (x),F_{{\mathbf k}}^{(\mp )} (x)]&=&0.
\end{eqnarray*}
The completeness condition is obvious:
\begin{equation}
\phi (x)=\int  \frac{d\mathbf{k}}{(2\pi )^{3} } \left(F_{{\mathbf k}}^{(+)} (x)i\int  d{\mathbf y} W_A [F_{\mathbf k}^{(+)*} (y),\phi (y)]-F_{{\mathbf k}}^{(-)} (x)i\int  d{\mathbf y} W_A [F_{{\mathbf k}}^{(-)*} (y),\phi (y)]\right).
\label{extf}
\end{equation}

In the theory of a charged scalar field, the canonical momenta are defined by equations $\pi^{\ast} (x) = D_t \phi (x)$ and $\pi (x) = (D_t \phi (x))^{\ast}$. The canonically conjugate variables satisfy
\begin{equation}
\{\phi(x),\pi(y)\}|_{x^0 = y^0} = \{\phi(x)^{\ast},\pi^{\ast}(y)\}|_{x^0 = y^0} = \delta(\mathbf{x} - \mathbf{y}), \label{PB6}
\end{equation}
while other pairs have the vanishing Poisson bracket.
The generalization of the corresponding relations of a free scalar field can be written as follows
\begin{eqnarray}
\int \frac{d\mathbf{k}}{(2\pi)^3} \left(
F^{(+)}_{\mathbf{k}}(x)F^{(+)\ast}_{\mathbf{k}}(y) - F^{(-)}_{\mathbf{k}}(x)F^{(-)\ast}_{\mathbf{k}}(y) \right)|_{x^0 = y^0}
 &=&0, \label{Fund7} \\
\int \frac{d\mathbf{k}}{(2\pi)^3} \left(
F^{(+)}_{\mathbf{k}}(x)D_t^{\ast } F^{(+)\ast }_{\mathbf{k}}(y) - F^{(-)}_{\mathbf{k}}(x)D_t^{\ast } F^{(-)\ast }_{\mathbf{k}}(y)
\right)|_{x^0 = y^0} &=& i \delta(\mathbf{x} - \mathbf{y}), \label{Fund8} \\
\int \frac{d\mathbf{k}}{(2\pi)^3} \left(
D_t F^{(+)}_{\mathbf{k}}(x)F^{(+)\ast }_{\mathbf{k}}(y) -
D_t F^{(-)}_{\mathbf{k}}(x) F^{(-)\ast }_{\mathbf{k}}(y)
\right)|_{x^0 = y^0} &=& -i \delta(\mathbf{x} - \mathbf{y}). \label{Fund9}
\end{eqnarray}
Equations (\ref{Fund7}) and (\ref{Fund8}) show that the completeness condition (\ref{extf}) holds for arbitrary functions evaluated at $x^0 = y^0$.

In conclusion we note that the zeroth component of vector potential can be removed by a gauge transformation, in which case $W_A = W$ and other relations and their proofs take the form more similar to the free case.

\subsection{Feynman propagator}

The decomposition of the Feynman propagator over the basis functions has the form
\begin{equation}
\Delta _{F} (x,y)=-i\int  \frac{d\mathbf{k}}{(2\pi )^{3} } \left(F_{{\mathbf k}}^{(+)} (x)F_{{\mathbf k}}^{(+)*} (y)\theta (x^{0} -y^{0} )+F_{{\mathbf k}}^{(-)} (x)F_{{\mathbf k}}^{(-)*} (y)\theta (-x^{0} +y^{0} )\right).
\label{proext}
\end{equation}
The use of Eqs.~(\ref{Fund7}) and (\ref{Fund9}) allows to verify by the direct calculation that
\begin{equation}
(D_{\mu } D^{\mu } +m^{2} )\Delta _{F} (x,\xi )=-\delta ^{4} (x-\xi ).
\label{44}
\end{equation}

\subsection{Superposition principle from Kirchhoff's integral theorem }

To derive Eq.~(\ref{31}), the identity (\ref{30}) was used.
After recapitulating the arguments used in
the proof of Eq.~(\ref{43}), we rewrite the divergence of
\begin{equation*}
\varphi \stackrel{\leftrightarrow}{D}_{\mu } \chi \equiv \varphi (D_{\mu } \chi )-(D_{\mu }^{*} \varphi )\chi ,
\end{equation*}
where $\varphi $ and $\chi $ are arbitrary functions, in the form
\begin{equation*}
\partial _{\mu } (\varphi \stackrel{\leftrightarrow}{D^{\mu }}  \chi )=\varphi (D_{\mu } D^{\mu } \chi )-(D^{*} _{\mu } D^{*\mu } \varphi )\chi .
\end{equation*}
Substituting $\Delta _{F} (x,\xi )$ and $\phi (\xi )$ in place of $\varphi $ and $\chi $, respectively, we obtain
\begin{equation}
\phi (\xi )\delta ^{4} (x-\xi )=\frac{\partial }{\partial \xi _{\mu } } \left( \Delta _{F} (x,\xi )(D_{\mu } \, \phi (\xi ))-(D_{\mu }^{*} \Delta _{F} (x,\xi ))\phi (\xi )\right).
\label{45}
\end{equation}

By choosing as the integration region a four-dimensional space with the variable $\xi ^{0} $ running in the interval $(y^{0} ,z^{0} )$, we find for $x^{0} \in (y^{0} ,z^{0} )$
\begin{eqnarray}
\phi (x) = \int d{\mathbf z}W_{A} [\Delta _{F} (x,z),\phi (z)] -\int d{\mathbf y}W_{A} [\Delta _{F} (x,y),\phi (y)].
\end{eqnarray}
In the opposite case of $x^{0} \notin (y^{0} ,z^{0} )$ the left-hand side vanishes.

\subsection{Superposition principle from the completeness condition}

The linearity of the evolution equation allows for the generalization of the superposition principle (\ref{35}) in the presence of an external electromagnetic field. The completeness condition leads to the following scheme:
\begin{equation}
\phi ^{(+)} (x)\theta (x^{0} -y^{0} )-\phi ^{(-)} (x)\theta (-x^{0} +y^{0} )=-\int d{\mathbf y}W_{A} [\Delta _{F} (x,y),\phi (y)]  .
\label{46}
\end{equation}
Under the integral sign, the derivative entering $W_A$ also generates the term
\begin{equation*}
\Delta(x,y) = -i \int  \frac{d\mathbf{k}}{(2\pi )^{3} } \left(
F_{\mathbf k}^{(+)} (x) F_{\mathbf k}^{(+)*} (y)
-     F_{\mathbf k}^{(-)} (x) F_{\mathbf k}^{(-)*} (y) \right)
\end{equation*}
multiplied by $\phi (y) \delta (x^{0} -y^{0})$.
In view of the relationship $x^{0} = y^{0}$ and Eq.~(\ref{Fund7}), this term vanishes. By calculating the derivative of Eq.~(\ref{46}) with respect to $y^{0} $, one can prove that the propagator obeys equation
\begin{equation}
(D_{\mu }^{*} D_{}^{\mu *} +m^{2} )\Delta _F (x,\xi )=-\delta (x-\xi ),
\label{47}
\end{equation}
where the differentiation is over $\xi $. This equation is equivalent to Eq.~(\ref{44}), where $D^{\mu } $ acts on $x$.

The superposition scheme for the retarded propagator is as follows
\begin{equation}
\phi (x)\theta (x^{0} -y^{0} )=-\int d{\mathbf y}
W_{A} [\Delta _{\mathrm{ret}} (x,y),\phi (y)] .
\label{48}
\end{equation}
This equation is the analog of Eq.~(\ref{17}). It can also be derived from Eq.~(\ref{45}).

To conclude, the superposition schemes for a free scalar field are fundamentally valid for a scalar complex field in an external electromagnetic field.

\section{Nonlinear field theory}
\renewcommand{\theequation}{V.\arabic{equation}}
\setcounter{equation}{0}

The superposition principle for secondary waves, which is the consequence of the GF method, should be distinguished from the superposition principle
as a manifestation of the linearity of the problem.
In linear theory, the wave is a source of secondary waves. 
In nonlinear theory, two sources of secondary waves exist: the wave itself plus a function $V^{\prime}(\phi)$. In both cases, secondary waves satisfy free linear wave equations, so the superposition principle applies to secondary waves universally.

\subsection{Superposition principle from Kirchhoff's integral theorem}

For a Lagrangian ${\rm {\mathcal L}}={\rm {\mathcal L}}_{{\rm free}} - V$, that contains a term $V=V(\phi )$ of a general form, the identity (\ref{30}) is modified as follows:
\begin{eqnarray}
\phi(\xi )\delta^{4} (\xi - x) &=& \Delta_{F} (x-\xi )
  ((\Box_{\xi } + m^{2} )
\phi(\xi )+ V'(\phi (\xi )))
- ((\Box_{\xi } + m^{2} )
\Delta_{F} (x-\xi ))\phi (\xi ) \nonumber \\
&=& \frac{\partial }{\partial \xi _{\mu}}
\left(
\Delta _{F} (x - \xi )
\frac{\stackrel{\leftrightarrow}{\partial }}{\partial \xi ^{\mu } } \phi (\xi )
\right)
+\Delta _{F} (x-\xi )V'(\phi (\xi )).
\end{eqnarray}
For $x^{0} \in (y^{0} ,z^{0} )$, this equation gives
\begin{equation}
\phi (x)=\phi _{0} (x)-\int d^{4} \xi  \Delta _{F} (x-\xi )V'(\phi (\xi )),
\label{49}
\end{equation}
where the integration over $\xi ^{0} $ runs over $\xi ^{0} \in (y^{0},z^{0})$ and the integral in $\boldsymbol\xi$ extends over all space. The field $\phi _{0} (x)$ is defined by the relation
\begin{equation}
\phi _{0} (x)=\int d{\mathbf z}W[\Delta _{F} (x-z),\phi (z)] -\int d{\mathbf y}W[\Delta _{F} (x-y),\phi (y)] .
\label{50}
\end{equation}
For $x^{0} \in (y^{0} ,z^{0} )$, $\phi _{0} (x)$ satisfies the free Klein-Gordon equation. If $x^{0} \notin (y^{0} ,z^{0} )$, then
\begin{equation}
0=\phi_{0}(x) - \int d^{4} \xi  \Delta _{F} (x-\xi ) V'(\phi (\xi )).
\label{51}
\end{equation}

In quantum field theory, Eq.~(\ref{49}) in the infinite limits $(y^{0} ,z^{0} )=(-\infty ,+\infty )$ is used in the development of perturbation theory. Unlike in the canonical formulation of the Fresnel superposition scheme, the integrand contains the nonlinear term $V'(\phi (\xi ))$ as an additional source of secondary waves and the integration spans the entire four-dimensional space.

Equations (\ref{49})  -  (\ref{51}) in nonlinear scalar field theory are analogous to Eqs.~(\ref{22}) - (\ref{24}) in the anharmonic oscillator problem.

The mass term of ${\rm {\mathcal L}}$ can be attributed either to
${\rm {\mathcal L}}_{{\rm free}}$ or to the potential $V$.
In the last case, ${\rm {\mathcal L}}_{{\rm free}}$ describes massless particles.
This might seem disadvantageous, because asymptotic states of ${\rm {\mathcal L}}$ are massive in general.
The positive feature is that the retarded Green's function of massless particles,
being localized on the light cone (see Eq.~(\ref{A1})),
ensures reduction of four-dimensional integrals in Eqs.~(\ref{49}) and (\ref{51}) to three-dimensional integrals
and transformation of integrals in Eq.~(\ref{50}) to surface integrals.

\subsection{Positive- and negative-frequency solutions}

Interacting fields can be decomposed into a sum of positive- and negative-frequency solutions only asymptotically.
In Sect. II.C.2, we demonstrated that
the straightforward generalization of the Fresnel superposition scheme to nonlinear dynamical systems
is possible and consistent; however,
its value is limited to only providing the definitions of positive- and negative-frequency solutions for arbitrary $t$. For the sake of completeness, we present here a field theoretical version of the nonlinear superposition scheme (\ref{26}):
\begin{eqnarray}
  \phi^{(+)}(x)\theta (  x^{0} - y^{0})
- \phi^{(-)}(x)\theta (- x^{0} + y^{0})
= &-&\int d\mathbf{y}W[\Delta _{F}(x - y),\phi (y)] \nonumber \\
  &+&\int d^4 \xi  \Delta _{F}(x - \xi)V'(\phi(\xi)),
\label{261}
\end{eqnarray}
where the integral over $ \xi^0 $ runs from $y^0$ to $x^0$.

The derivative over $y^{0}$ leads to the relation $\phi (t)=\phi^{(+)} (t)+\phi^{(-)} (t)$ and Eq.~(\ref{29}). The difference in Eq.~(\ref{261}) at two different time points leads to Eqs.~(\ref{49})  -  (\ref{51}). Equation (\ref{261}) ensures that
$\phi^{(\pm )} (x)$ is a linear superposition in $\mathbf{k}$ of the basis functions $f^{(\pm )}_{\mathbf k} (x)$ at $t\to \pm \infty $.

The calculation of the first derivative of Eq.~(\ref{261}) in $x^{0} $ yields a superposition scheme for the canonical momentum:
\begin{eqnarray}
  \pi^{(+)}(x)\theta (  x^{0} - y^{0})
- \pi^{(-)}(x)\theta (- x^{0} + y^{0})
=& -&\int d\mathbf{y}W[\Delta _{F}(x - y),\pi (y)] \nonumber \\
  &+&\int d^4\xi       \Delta _{F}(x - \xi)V''(\phi(\xi))\pi(\xi),
\label{262}
\end{eqnarray}
where the integral over $\xi^0$ runs from $y^0$ to $x^0$.



\section{Conclusions}
\renewcommand{\theequation}{6.\arabic{equation}}
\setcounter{equation}{0}

The evolution of the ideas underlying the Huygens-Fresnel superposition principle from geometrical and wave optics to the theory of interacting fields is highly instructive:

In geometrical optics, a \textit{wave front} refers to the two-dimensional surface that defines the farthest extent to which the wave has arrived after a certain period of time. Huygens' principle (1678), based on the Fermat principle, allows for the determination of how the wave front is propagating.

In wave optics, the term wave front has no strict definition. Instead, the term \textit{wave surface} is used. The wave surface is the two-dimensional surface on which the phase of the wave is constant.
A.-J. Fresnel proposed the principle of superposition (1816), which details the wave process. A wave is a
result of interference of secondary waves
emitted at an earlier time. At any fixed point, it is determined by the phase and amplitude at a wave surface corresponding to a preceding instant of time. The wave surface in the past can be chosen arbitrarily. The superposition principle anticipates informal content of the GF method (1828).

Kirchhoff's integral theorem (1883) is a dynamic, four-dimensional extension of Green's third identity of the static potential theory. More than half a century separates this theorem from Green's major work \cite{GGreen1828}, which introduced
the basic concepts of the GF method.
\footnote{In 1839 year G. Green came closely to the notion of the four-dimensional Green's function. The value of the GF method in quantum field theory is highly appreciated. \cite{Schw93}}
Kirchhoff's integral theorem provides a mathematical proof of the superposition principle, clarifying and quantifying it.

First, the theorem demonstrates that the amplitude of the secondary waves is determined by the Wronskian of the Green's function and the field at a previous time.

Second, the wave surfaces are not highlighted; this is perfectly consistent with the fact that they are not necessarily observable (in the massive theory, e.g., the speed of a wave surface of a plane wave is always greater than the speed of light). The surface must be closed and contain the point at which the wave is calculated; otherwise, it can be arbitrarily chosen. Outside the closed surface, the interference of the secondary waves is strictly destructive: for any exterior point, the calculation of the surface integral yields zero.

The reasoning used in the proof
can be regarded as a standard piece of the GF method;
it is of high generality, goes beyond the problem of propagation of electromagnetic waves and allows for an understanding of how the superposition principle should be modified in the theory of interacting fields. Note the most significant modifications:

i)~~According to the Huygens-Fresnel superposition principle, a wave at a given point is expressed as a superposition of secondary waves emitted from centers located on a two-dimensional surface. This property arises only in massless theories, including the theory of electromagnetic fields, where the group and phase velocities coincide with the speed of light, which is the necessary condition for the integral over time delay and advance be not available in Eq.~(\ref{40}). Kirchhoff's integral theorem for massive particles, Eq.~(\ref{38}), states that a wave is determined by its values on a closed surface at all times. The physical interpretation of this fact is quite transparent. The Fourier expansion of a massive field contains components of various momenta corresponding to various group and phase velocities, which leads to a spread in time lags. As a result, the two-dimensional integral over the sources of secondary waves is transformed into a three-dimensional integral.

ii)~~In the nonlinear theory,
there is a need for a more extensive modification of the superposition scheme. In addition to the wave itself,
a nonlinear function of the field $V'(\phi (\xi ))$ becomes the source of secondary waves. The summation runs over distributed sources: from a two-dimensional surface in theories with massless particles to a two-dimensional surface and the time axis in theories with massive particles and the entirety of four-dimensional space. This type of representation holds for both local nonlinear and nonlocal theories.

We see that after each modification, the effectiveness of the superposition principle weakens. In the most general nonlinear case, the modified principle certainly does not promise fast 
results. To determine the field, it is necessary to calculate a four-dimensional integral in a self-consistent manner. In linear theories, the superposition principle solves the evolution problem, but in nonlinear theories, it only offers a different formulation of the problem. Nevertheless, relations of this type are still useful when searching for solutions within the framework of perturbation theory, when the non-linearity is small. In other cases, the solutions found using other techniques can be checked. The four-dimensional representation given by Eq.~(\ref{49}) is a consequence of Kirchhoff's integral theorem, but in quantum field theory, it is typically derived directly from the properties of Green's functions.

In the context of a field theory,
the original form of
the superposition principle only has
heuristic value.
The superposition schemes for the secondary waves that are used to solve specific problems are unified by Kirchhoff's integral theorem, which exploits the properties of the Wronskian of the Green's functions and solutions of the wave equation under consideration. The spectrum of such problems is quite comprehensive: from the harmonic oscillator to scalar electrodynamics and nonlinear field theories.

In addition to the use of the GF method, which has found a variety of applications in quantum field theory, Kirchhoff's theorem has a wider range of corollaries. Equation (\ref{38}), which represents one version of Kirchhoff's theorem, does not arise in quantum field theory because of the boundary conditions, which are atypical of a scattering problem. However, the superposition scheme represented by Eq. (\ref{35}), which is not based on Kirchhoff's theorem, is not sufficiently general because it does not extend to theories with interaction.

The statement of Kirchhoff's integral theorem depends on the asymptotic conditions imposed on the Green's function.
In the main part of the paper, because we were interested in the place of this remarkable principle and well-known theorem in quantum field theory, we applied the Feynman asymptotic conditions almost universally.

\begin{acknowledgments}
This work was supported in part by RFBR Grant No.~16-02-01104 and Grant~No.~HLP-2015-18~of Heisenberg-Landau~Program.
\end{acknowledgments}

\appendix

\renewcommand{\theequation}{A.\arabic{equation}}
\renewcommand{\thesection}{}
\setcounter{equation}{0}

\section{Kirchhoff's integral theorem and its vector extensions 
with the retarded Green's function}
\renewcommand{\theequation}{A.\arabic{equation}}
\setcounter{equation}{0}

In the main sections of the paper, emphasis is placed on the Feynman asymptotic conditions,
which play a special role in quantum field theory.
Here, we formulate Kirchhoff's integral theorem and its vectorial generalizations for the retarded Green's function.

\renewcommand{\thesubsection}{A.1}
\subsection{Free massless scalar field}

Equation~(\ref{32}) is essentially the third Green's identity for time-dependent solutions of the wave equation;
its proof is outlined in Sect. III.A. As noted earlier, Eq.~(\ref{32}) holds for any Green's function.

The retarded Green's function in the coordinate representation has the following form (see e.g., \cite{Bjor65} Appendix B)
\begin{eqnarray} \label{A1}
\Delta _{\mathrm{ret}}(t,{\mathbf x}) &=&\int \frac{d^{4}q}{(2\pi )^{4}}e^{-iqx}\frac{1}{q^{2}+i0\mathrm{sgn}(q_{0})} \nonumber \\
&=&-\frac{1}{4\pi |\mathbf{x}|}\left( \delta (|\mathbf{x}|-t)-\delta (|\mathbf{x}|+t)\right) \theta (t).
\end{eqnarray}
The product of generalized functions of a single variable is not defined. The propagator depends on four space-time coordinates.
Generalized functions of four variables allow for products of up to four generalized functions of one variable,
provided their arguments are independent.
$\Delta _{\mathrm{ret}}(t,{\mathbf x})$ is thus a well-defined generalized function.

$\Delta _{\mathrm{ret}}(t,{\mathbf x})$ is localized on the upper half of the light cone $t^2 - {\mathbf x}^2 = 0$.
Substituting (\ref{A1}) in place of $\Delta _{F} (t,{\mathbf x})$ in Eq.~(\ref{38}), one arrives at the original Kirchhoff representation \cite{Kirch1883,Born99}
\begin{eqnarray}
\phi _{0} (t,{\mathbf x})\theta ({\mathbf x}\in \Omega _{3} ) =
\frac{1}{4\pi } \int _{\partial \Omega _{3} }dS_{{\boldsymbol\xi }}^{\alpha }  \left[
- \frac{1}{\rho } \frac{\partial }{\partial \xi ^{\alpha } } \right.
\phi _{0} (t - \rho/c ,{\boldsymbol\xi })
+\left(\frac{\partial }{\partial \xi ^{\alpha } } \frac{1}{\rho } \right)
\phi _{0} (t - \rho/c ,{\boldsymbol\xi }) && \nonumber \\
- \left. \frac{1}{\rho } \frac{\partial \rho }{\partial \xi ^{\alpha } }  \frac{\partial }{c \partial t}
\phi _{0} (t - \rho/c ,{\boldsymbol\xi })
\right],&&
\label{401}
\end{eqnarray}
where $\rho =\left|{\boldsymbol\xi }-{\mathbf x}\right|$ and where
the differentiation in $\xi ^{\alpha } $ does not affect $\rho $ in the first term.
The dependence on the speed of light $c$ is here made explicit.
The wave $\phi _{0} (t,{\mathbf x})$ at ${\mathbf x}\in \Omega _{3}$ is determined by its values
on the closed surface $\partial \Omega _{3}$ considering the delay $\rho/c$.
Linear second-order hyperbolic partial differential equations possessing this property are well-studied from a mathematical point of view \cite{Cour1962,Gunt1991,Duis1992}.

\renewcommand{\thesubsection}{A.2}
\subsection{Monochromatic electromagnetic fields with sources}

A generalization of Kirchhoff's integral theorem, which takes into account vectorial character of the electromagnetic
field and the electromagnetic currents, is obtained by von Ignatowsky. \cite{Igna1907}
First, however, we consider a generalization of von Helmholtz's theorem, following Stratton and Chu \cite{Stra1939}.

Most methods used in Sect.~III.C.2 for a free monochromatic scalar field
 apply to a monochromatic electromagnetic field with sources after some slight modifications.
We replace scalar field $\phi _{0}$ by the electromagnetic
field tensor $F^{\mu \nu }=\partial ^{\nu }A^{\mu }-\partial ^{\mu }A^{\nu }$.
In the Lorentz gauge $\partial _{\mu }A^{\mu }=0$,
the evolution equations $\partial _{\nu }F^{\mu \nu }=j^{\mu }$
become $\square A^{\mu }=j^{\mu }$,
where $j^{\mu }$ is the electromagnetic current.
It is assumed that the fields are harmonic and that all quantities contain
a factor $\exp (-i\omega t)$, so that $\partial ^{0}j^{\mu }=-i\omega j^{\mu }$,
$(\omega ^{2}+\triangle )A^{\mu }=-j^{\mu }$, and
\begin{equation} \notag 
(\omega ^{2}+\triangle )F^{\mu \nu }=-\partial ^{\nu }j^{\mu }+\partial ^{\mu
}j^{\nu }.
\end{equation}
Since the right-hand side
of this equation is different from zero, the analogue of Eq.~(\ref{G2I}) takes a more complicated form:
\begin{eqnarray}
F^{\mu \nu }(\omega ,{\boldsymbol{\xi }})\delta ({\boldsymbol{\xi }}-{%
\mathbf{x}}) &=&\Delta _{\mathrm{ret}}(\omega ,{\mathbf{x}}-{\boldsymbol{\xi
}})(-\partial ^{\nu }j^{\mu }(\omega ,{\boldsymbol{\xi }})+\partial ^{\mu
}j^{\nu }(\omega ,{\boldsymbol{\xi }}))  \notag \\
&&-\frac{\partial }{\partial \xi ^{\alpha }}\left( \Delta _{\mathrm{ret}%
}(\omega ,{\mathbf{x}}-{\boldsymbol{\xi }})\frac{\overset{\leftrightarrow }{%
\partial }}{\partial \xi ^{\alpha }}F^{\mu \nu }(\omega ,{\boldsymbol{\xi }}%
)\right) .  \label{30 bis}
\end{eqnarray}%
The sum in $\alpha$ runs from 1 to 3, while $\mu,\nu = 0, 1, 2, 3$.
By integrating over a three-dimensional region $\Omega _{3}$, one gets
\begin{eqnarray}
F^{\mu \nu }(\omega ,{\mathbf{x}}) \theta ({\mathbf x}\in \Omega _{3} ) &=&
\int_{\Omega _{3}}d{\boldsymbol{\xi }}\Delta _{\mathrm{ret}}(\omega ,{\mathbf{x}}-{\boldsymbol{\xi }})
(-\partial^{\nu }j^{\mu }(\omega ,{\boldsymbol{\xi }})+\partial ^{\mu }j^{\nu }(\omega
,{\boldsymbol{\xi }}))  \notag \\
&&-\int_{\partial \Omega _{3}}dS_{\xi }^{\beta }\Delta _{\mathrm{ret}%
}(\omega ,{\mathbf{x}}-{\boldsymbol{\xi }})\frac{\overleftrightarrow{%
\partial }}{\partial \xi ^{\beta }}F^{\mu \nu }(\omega ,{\boldsymbol{\xi }}).
\label{31 bisbis}
\end{eqnarray}%
The dependence on the derivatives of $F^{\mu \nu }$ can be eliminated. \cite{Stra1939}

\renewcommand{\thesubsection}{A.3}
\subsection{Non-monochromatic electromagnetic fields with sources}

In the presence of external currents, electromagnetic fields satisfy the identity
\begin{eqnarray}
F^{\mu \nu }(\xi )\delta ^{4}(\xi -x) &=&
\Delta _{\mathrm{ret}}(x-\xi)(-\partial ^{\nu }j^{\mu }(\xi )+\partial ^{\mu }j^{\nu }(\xi ))  \notag \\
&&-\frac{\partial }{\partial \xi_{\sigma }}\left( \Delta _{\mathrm{ret}}(x-\xi
)\frac{\overset{\leftrightarrow }{\partial }}{\partial \xi ^{\sigma }}F^{\mu
\nu }(\xi )\right) .  \label{30 time}
\end{eqnarray}
The sum in $\sigma$ runs from 0 to 3.
By taking the integral over a four-dimensional region $\Omega $, we obtain
\begin{eqnarray}
F^{\mu \nu }(x)\theta (x \in \Omega ) &=& \int_{\Omega }d^{4}\xi \Delta _{%
\mathrm{ret}}(x-\xi )(-\partial ^{\nu }j^{\mu }(\xi )+\partial ^{\mu }j^{\nu
}(\xi ))  \notag \\
&&-\int_{\partial \Omega }^{{}}dS_{\xi }^{\sigma }\left( \Delta _{\mathrm{ret}%
}(x-\xi )\frac{\overset{\leftrightarrow }{\partial }}{\partial \xi ^{\sigma}}%
F^{\mu \nu }(\xi )\right).  \label{32 bis}
\end{eqnarray}
The representation becomes linear in $F^{\mu \nu }$ after replacing $j^{\mu }$ with $\partial _{\nu }F^{\mu \nu }$.
The field derivatives are assumed to be smooth.

Equation~(\ref{32 bis}) can be simplified by choosing $\Omega $
to be an infinite cylinder,
$\Omega = \mathbb{R}^1 \otimes \Omega _{3} $,
whose cross section is a three-dimensional space-like region $\Omega_{3}$.
With the use Eq.~(\ref{A1}), the integration over the time coordinate gives \cite{Igna1907}
\begin{eqnarray}
F^{\mu \nu }(t,{\mathbf{x}})\theta ({\mathbf{x}} \in \Omega _{3})&=&
-\frac{1}{4\pi }\int_{\Omega _{3}}d{\boldsymbol{\xi }}\frac{1}{\rho }
(-\frac{\partial}{\partial \xi_{\nu}}j^{\mu }(t-\rho,{\boldsymbol{\xi }})+\frac{\partial}{\partial \xi_{\mu}}j^{\nu}(t-\rho,{\boldsymbol{\xi }})) \nonumber \\
&&+\frac{1}{4\pi }\int_{\partial \Omega _{3}}dS_{{\boldsymbol{\xi }}%
}^{\alpha }\left[ -\frac{1}{\rho }\frac{\partial }{\partial \xi ^{\alpha }}%
\right. F^{\mu \nu }(t-\rho,{\boldsymbol{\xi }}) \nonumber \\
&&+ \left( \frac{\partial }{%
\partial \xi ^{\alpha }}\frac{1}{\rho }\right) F^{\mu \nu }(t-\rho,{%
\boldsymbol{\xi }})-
\left. \frac{1}{\rho }\frac{\partial \rho }{\partial \xi
^{\alpha }}\frac{\partial }{\partial t}F^{\mu \nu }(t-\rho,{\boldsymbol{%
\xi }})\right], \label{a8}
\end{eqnarray}%
where $\rho =\left\vert {\boldsymbol{\xi }}-{\mathbf{x}}\right\vert$.
In the first two lines, the differentiation
with respect to $\xi ^{\alpha }$  does not apply to $\rho $. 

Kirchhoff's integral theorem (\ref{401}) and Eq.~(\ref{31 bisbis}) extend von
Helmholtz's theorem (\ref{37}) 
in different directions.
Equation (\ref{a8}) constitutes, from one hand, the generalization of
Kirchhoff's integral theorem by taking into account the vectorial character of electromagnetic field
and including the effect of electromagnetic currents and, from
other hand, the generalization of Eq.~(\ref{31 bisbis})
by going beyond the monochromatic field assumption.


\end{document}